\documentclass[twocolumn,aps,showpacs,prlprintnumbers,fleqn,superscriptaddress]{revtex4}
\usepackage{epsfig}
\usepackage{graphicx}
\usepackage{amsfonts}
\usepackage{subfigure}
\usepackage[dvips]{color}
\newcommand{\ig}{\includegraphics}
\newcommand{\ct}{\cite}
\newcommand{\bi}{\bibitem}

\newcommand{\ket}{\rangle}
\newcommand{\non}{\nonumber}

\newcommand{\be}{\begin{equation}}
\newcommand{\ee}{\end{equation}}
\newcommand{\ba}{\begin{eqnarray}}
\newcommand{\ea}{\end{eqnarray}}

\begin{document}

\title{Quenching through Dirac and semi-Dirac points in optical Lattices:
Kibble-Zurek scaling for anisotropic Quantum-Critical systems}

\author{Amit Dutta}
\email{dutta@iitk.ac.in}
\affiliation{Department of Physics, Indian Institute of Technology, Kanpur 208 016, India}
\author{R. R. P. Singh}
\affiliation{Department of Physics, University of California, Davis, CA95616, USA}
\email{singh@physics.ucdavies.edu}
\author{Uma Divakaran}
\affiliation{Department of Physics, Indian Institute of Technology, Kanpur 208 016, India}
\email{udiva@iitk.ac.in}

\date{\today}
\begin{abstract}
We propose that Kibble-Zurek scaling can be studied
in optical lattices by creating geometries that support, Dirac, Semi-Dirac and
Quadratic Band Crossings. On a Honeycomb lattice with fermions, as a staggered on-site potential
is varied through zero, the system crosses the gapless Dirac points, and we show that
the density of defects created scales as $1/\tau$, where $\tau$ is the 
inverse rate of change 
of the potential, in agreement with the Kibble-Zurek relation.
We generalize the result for a passage through a semi-Dirac point in $d$ dimensions,
in which spectrum is  linear in $m$ parallel directions  and  quadratic in 
rest of the perpendicular $(d-m)$ directions. We find that the defect density is
 given by $  1 /{\tau^{m\nu_{||}z_{||}+(d-m)\nu_{\perp}z_{\perp}}}$
where $\nu_{||}, z_{||}$ and $\nu_{\perp},z_{\perp}$ are the dynamical exponents and 
the correlation length exponents along the parallel and perpendicular 
directions, respectively. The scaling relations are also generalized to the case 
of non-linear quenching.

\end{abstract}

\maketitle
The Kibble-Zurek (KZ) scaling \ct{kibble76,zurek96,zurek05,polkovnikov05,damski05,dziarmaga05} of defect density in the final state of a quantum many-body system following a slow passage across a quantum critical point, has been an exciting
area of recent research. The KZ argument predicts that the scaling of the defect density is universal and is given as $ n \sim  1/\tau^{\nu d/(nu z+ 1)} $  where $\tau$ is the inverse rate
of change of a parameter, $d$ is the spatial dimension and $\nu$ and $z$ are the correlation length and dynamical exponents, respectively, associated with the quantum
critical point \ct{sachdev99,dutta96} across which the system is swept. Following
the initial predictions, a plethora of theoretical studies have been carried out 
 \ct{polkovnikov07,levitov06,mukherjee07,
sengupta08,sen08,santoro08,caneva08,
divakaran09,barankov08} to explore the defect generation and the entropy production using different
quenching schemes across critical points \ct{levitov06}, quantum multicritical points \ct{divakaran09}, gapless phases \ct{sengupta08}, along gapless lines \ct{santoro08}, etc. The possibility of experimental observations in a spin-1 Bose condensate \ct{sadler06} and also on ions trapped in optical lattices 
\ct{duan03,bloch07} has provided a tremendous boost to the related theoretical studies.

Here, we propose that Kibble-Zurek scaling can be studied in optical lattices with
fermionic atoms by creating geometries that support Dirac, semi-Dirac and Quadratic
Band Crossings. For example, a Honeycomb lattice consists of two 
interpenetrating triangular lattices. If the two sublattices are
controlled separately, one can create a staggered on-site potential,
which creates a gap in the spectrum \cite{graphene1,semenoff08}.
We will call the Honeycomb lattice with a staggered potential
a gapped Graphene Hamiltonian in analogy with Graphene \cite{graphene1}.
The system can be loaded with
atoms when one of the sublattices has a much lower energy than the
other. As the staggered potential is varied through
zero, the system will cross through gapless Dirac point, and the creation 
of defect density can be studied as a function of the rate of change of 
potential. Similarly, two interpenetrating square-lattices can lead
to Quadratic Band Crossing \cite{kivelson09} and 
either a deformed Honeycomb system \cite{giles1,giles2}, or a 3-band system can lead to semi-Dirac
points\cite{banerjee09}, where the spectrum is linear along one axis and quadratic
along another.
Anisotropic quantum critical points associated with spectra that
are linear in some directions and quadratic in others arise in
systems as diverse as semiconductor hetero-structures \cite{pickett}
and He$^3$ \cite{volovik01}.
We will calculate a generalized Kibble-Zurek scaling for such
anisotropic Quantum critical systems.

Many  previous theoretical studies are on one-dimensional quantum spin
systems some of which can be exactly solved via Jordan-Wigner transformation \ct{lieb61}. 
On the other hand, the quenching dynamics of non-integrable spin chains have been studied using adiabatic perturbation theory \ct{polkovnikov05} or  exact diagonalization and time-dependent density matrix renormalization techniques \ct{santoro08,caneva08}. Both the methods have been very successful in yielding exciting results for the scaling of defect  density following a 
slow quantum quench in pure as well as random systems \ct{caneva08}.  
There are also a few results 
for the quenching dynamics of higher dimensional systems, e.g., in 
ref. \ct{sengupta08}  the exact solvability of a two-dimensional Kitaev model \ct{kitaev06} 
was utilized.

In this paper, we propose a kind of time-dependent 
perturbation that would be very
difficult to achieve in solid state systems but should be possible
in optical lattices. Because the underlying geometry in optical lattices
is controlled by
lasers, one should be able to change them in such a way that the system
passes through special band crossings. The salient feature of our
proposal on Honeycomb and other lattices is that 
the time-dependent perturbation preserves crystal momentum. 
Thus, as long as the interactions
between particles are weak, the system factorizes into independent momentum sectors.
For each value of the momentum, there
is a probability that the system can undergo transition from the lower to the upper
eigenstate. This can be analyzed using the
Landau-Zener transition formula \ct{landau} to provide an exact result for the defect
creation in the quenching process. Integration over all momentum leads
to the desired KZ scaling relations. Furthermore, quenching through
the semi-Dirac point yields a more general form of the KZ scaling for 
spatially anisotropic quantum critical points.
In principle, Dirac-like band crossings are of tremendous interest in themselves, 
in contexts such as topological insulators.\cite{kane08,zhang08}
But, whether one can find a perturbation that drives the system through
such a crossing will have to be examined in individual cases.

\begin{figure}
\ig[height=1.5in,width=2.0in,angle=0]{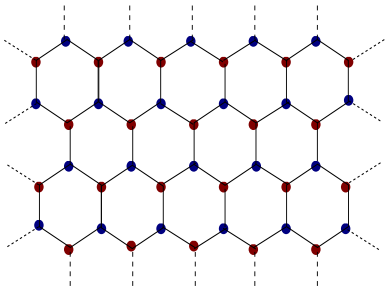}
\caption{A schematic picture of a honeycomb lattice consisting of two interpenetrating
triangular lattices shown in different colors. In cold-atoms, this
system can be controlled by two sets of lasers.}
\label{honeycomb}
\end{figure}

We focus primarily on a tight-binding model of fermions 
moving on a Honeycomb lattice. This lattice
has two inequivalent sites per unit cell (see Fig.~1) and hence the system is described
in the momentum space in terms of de-coupled $2 \times 2$ matrices \ct{graphene1}. Assuming the system
to be half-filled, the honeycomb lattice of the tight-binding Hamiltonian is critical
at two inequivalent points of the Brillouin zone and there is Dirac cone of excitations
with linear dispersion around this point. If a staggered on-site potential (called
the mass term) which differentiates between two inequivalent sites is added to the tight binding Hamiltonian, a gap is created  in the spectrum \ct{semenoff08}.
We assume that the magnitude of the mass term (denoted by $\mu$) is quenched linearly from $-\infty$
to $-\infty$ at a linear rate $ \mu=t/\tau$ so that the system crosses the gapless
point $\mu=0$ at a time $t=0$. When $\mu=-\infty$, only one sublattice of the HC lattice
will be occupied which amounts  to saying that for every Fourier mode $k$, one of the basis 
states is occupied. If the mass term is slowly changed, the state would follow adiabatically
until the Dirac point at $t=0$ where the spectrum is gapless and thereafter system no longer follows the ground state adiabatically. We here show that the density of defect thus generated is in
complete agreement with the KZ prediction. We then generalize to the case of quenching
through a $(d,m)$ semi-Dirac point \ct{banerjee09} at which the spectrum is linear in $m$-directions
and  quadratic in rest of the $(d-m)$-directions. Exploiting again 
the $ 2 \times 2$ form of the resulting Hamiltonian and the  Landau-Zener transition formula we arrive at a generalized KZ scaling form  $  1 /{\tau^{m\nu_{||}z_{||}+(d-m)\nu_{\perp}z_{\perp}}}$
where $\nu_{||}, z_{||}$ and $\nu_{\perp},z_{\perp}$ are the dynamical exponent and the correlation length exponents along the parallel directions and perpendicular directions,
respectively. The same  scaling relation is  also expected to be  valid for quenching through a $(d,m)$ quantum Lifhshitz point \ct{dutta97}.

Let us start from the general Hamiltonian of fermions moving on a honeycomb lattice
\be
H=\sum_{k=0}^{\pi}\psi_k^\dag H_k \psi_k
\ee
where $H_k$ is the reduced $2 \times 2$ matrix $(H_k)_{11} = (H_k)_{22}=0$ and $(H_k)_{12}= (H_k)_{12}^{*}= -t[e^{ik_xa} +2 e^{-ik_xa/2} \cos (k_y \sqrt{3}a/{2})$, where $t$ is the hopping term and $a$ is the lattice spacing. In
the presence of a mass term that arises due to a staggered on-site potential $\mu$ \ct{semenoff08}, the
reduced Hamiltonian takes the form
\ba 
\tilde{H_k}=\left[ \begin{array}{cc} \mu  &  \Delta\\
\Delta^{*} & -\mu \end{array}\right]\label{ham1}  
 \ea
where  $\Delta(k)=(H_k)_{12}$ as defined before. The off-diagonal term can be
simplified by expanding around the Dirac point $(0, \frac {4\pi}{3\sqrt{3}a})$ as
$\Delta(k)= -t \left[e^{iq_x a} + 2 e^{-iq_x a/2}  \cos ({(\frac {4\pi} {3\sqrt{3}a}+q_y) \frac {\sqrt{3}a}{2})}\right]$ where $k_x=q_x$ and $k_y = \frac {4\pi} {3\sqrt{3}a}+q_y$. In
the limit of $q_x, q_y \to 0$, we arrive at a simpler form of Eq.~ \ref{ham1} 
\ba 
\tilde{H_k}=\left[ \begin{array}{cc} \mu  &  v_f(q_y-iq_x)\\
 v_f(q_y+iq_x) & -\mu \end{array}\right]\label{ham2}  
 \ea
where $v_f = 3ta/2$.

During the quenching $\mu=t/\tau$,
the general state vector at an instant $t$ can be written as $|\psi_k(t)\ket=C_{1k}(t)| {1_k}\ket
+C_{2k}(t)| {2_k}\ket$ where the basis vectors are  $|{1_k}\ket$ and $ |{2_k}\ket$. With the initial condition $|C_{1k}(t\to-\infty)|^2=1$, the 
non-adiabatic transition probability due to the passage through the Dirac point
is given as $p_k=|C_{1k}(t\to+\infty)|^2$. Using the Hamiltonian (\ref{ham1}), the time evolution of the system is 
described in terms of the Schr\"odinger equations 

\ba
i\frac{\partial}{\partial t}C_{1k}(t)&=& \frac t {\tau}  C_{1k}(t) + \Delta(k)C_{2k}(t)
\non\\
i\frac{\partial}{\partial t}C_{2k}(t)&=&-  \frac t {\tau}C_{2,k}(t) +  \Delta^{*}(k) C_{1k}(t).
\label{schr1}
\ea
The above equations describes   two-time dependent levels 
$\pm \sqrt{(t/\tau)^2 + \Delta^2}$ approaching each other and there is an avoided level crossing at $t=0$. The non-adiabatic transition probability can be calculated using
the Landau-Zener transition formula \ct{landau} for each Fourier mode $k$ given
by $ p_k= \exp (-\pi |\Delta|^2 \tau)$. We therefore obtain the density of defects integrating over the modes $n=\int p_k d^2k$. The non-adiabatic transition becomes
prominent only in the vicinity of the Dirac point where the gap closes, we  therefore use the form of  $\Delta$ valid close to the Dirac point given in Eq.~(\ref {ham2}) and
extend the limits of integration from $-\infty$ to $+\infty$. The defect density
therefore scales as
\be
n= \int_{-\infty}^{+\infty} dq_xdq_y\exp(-\pi\tau v_f^2(q_x^2+q_y^2)) \sim \frac 1 {\tau}
\ee
Noting that the spectrum close to the Dirac point is of the form 
$\pm \sqrt{(\mu^2 + v_f^2|q|^2)}$, we find that the minimum gap vanishes as $\mu$ 
yielding the critical exponents $\nu z=1$. On the other hand, the linear dispersion
at the Dirac point ($\mu=0$) yield the dynamical exponent $z=1$.
The $1/\tau$-scaling of the defect density therefore is in complete agreement with the
Kibble-Zurek prediction $n\sim 1/\tau^{\nu d/\nu z+ 1}$. The
defect density obtained via numerical integration
is shown in Fig.~2.

\begin{figure}
\ig[height=2.5in,angle=-90]{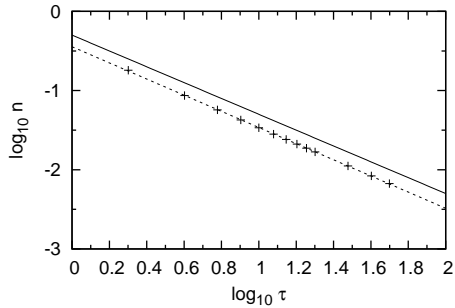}
\caption{The density of defect generated due to quenching through a two-dimensional
Dirac point, as in Graphene, obtained through the numerical integration of the
Schr\"odinger equations \ref{schr1}. The data clearly shows a $1/\tau$-decay of
the defect density which is indicated with a straight line of slope -1. }
\label{dirac}
\end{figure}

In the optical lattices, the defect density can be measured through the
total occupation of the upper band, or in more detail, by the momentum distribution
function, $n(k)$. In our two-band Graphene system, $n(k)$ in the lower-band
would develop a hole near the Dirac points, whose size would grow with
the rate of quenching. Conversely, in the upper-band the particles will be concentrated
close to the Dirac points. The resulting momentum
distribution would be highly unlike a thermal broadening. A picture of
the momentum distribution function in the upper band is shown in an extended 
Brillouin zone in Fig.~3.

\begin{figure}
\ig[height=2.5in,width=3.0in,angle=0]{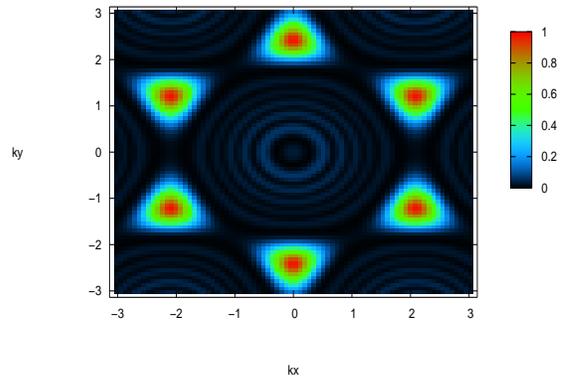}
\caption{Momentum distribution of particles created in the upper band is shown
in an extended Brillouin zone. The centers of the six bright regions form
the corners of the first Brillouin Zone and house the
two inequivalent Dirac points.}
\end{figure}

We shall now concentrate on the quenching through the semi-Dirac point which provides
a more general situation of the quenching dynamics. The Hamiltonian close to the
semi-Dirac point with a mass term can be put in the form \ct{banerjee09}
\ba 
\tilde{H_k}=\left[ \begin{array}{cc} \mu  &  iv_f |q_y|+ q_x^2/2m\\
 -iv_f |q_y|+ q_x^2/2m & -\mu \end{array}\right]\label{ham2}  
 \ea
If the mass term $\mu$ is quenched as $t/\tau$ and the system 
crosses the semi-Dirac point, the defect density scales as
\be
n=\int_{-\infty}^{\infty}dq_x dq_y \exp(-\pi \tau (v_f^2q_y^2+ \frac {q_x^4}{(2m)^2}))
\sim \frac 1 {\tau^{3/4}}\label{defect2}
\ee
The corresponding numerical result is shown in Fig.~2. Generalizing to three-dimesnions
\ct{volovik01}, where the off-diagonal term gets modified to $\Delta_q =iv_f |q_{||}|+ q_{\perp}^2/2m$
with  ${\vec q_{\perp}}^2=q_x^2+q_y^2$, the scaling of
 the defect
density is given by

\be
n=\int_{-\infty}^{\infty}dq_{||} d{\vec q_{\perp}} \exp(-\pi \tau (v_f^2q_{||}^2+ \frac {|q_{\perp}|^4}{(2m)^2}))
\sim \frac 1 {\tau}\label{defect3}
\ee
The scaling relations derived in Eqs.~(\ref{defect2}) and (\ref{defect3}) lead to a very
interesting generic scaling relation of the defect density  for quenching through a $(d,m)$
anisotropic quantum critical point given by the modified KZ form
\be
n\sim \tau^{-\left(\frac {m\nu_{||}}{\nu_{||}z_{||}+1}+\frac{(d-m)\nu_{\perp}}{\nu_{\perp}z_{\perp}
+1)}\right)},
\label{defect4}
\ee
where the critical exponents $(\nu_{||},z_{||})$ and $(\nu_{\perp},z_{\perp})$ are conjugate to parallel and perpendicular directions, respectively.
The case with $m=0$ refers to the quenching through a  band-crossing point with quadratic spectrum in
$d$-dimensions
while $m=d$ is the result for quenching through a Dirac point.
\begin{figure}
\ig[height=2.5in,angle=-90]{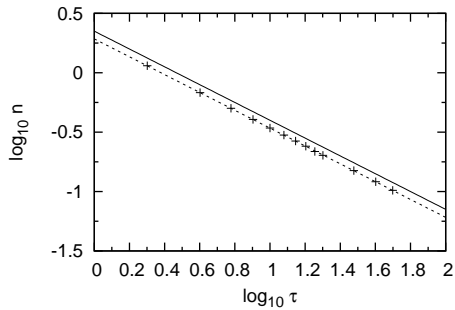}
\caption{The density of defect generated due to quenching through a two-dimensional
semi-Dirac point obtained through the numerical integration. The  $\tau^{-3/4}$-decay of
the defect density is indicated with a straight line of slope -3/4. }
\label{semi-dirac}
\end{figure}

We shall generalize the modified KZ scaling (\ref{defect4}) to the case of non-linear variation of the mass parameter $\mu=|t/\tau|^{\alpha}sgn(t)$. 
The situation is simpler as the spectrum is gapless for $\mu=0$ (i.e., $t=0$)
\ct{sen08}. Following refs. \ct{sen08,barankov08}, it is easy to show that
 that the 
probablity of excitation for the mode $k$, $p_k = |{\tilde C_{1k}(t\to \infty)}|^2$ must be
a function of $|\Delta(k)|^2\tau^{{2\alpha}/{\alpha+1}}$ where the initial condition $|{\tilde C_{1k}(t\to -\infty)}|^2=1$. Expanding around the
semi-Dirac point and extending the limits of integration form $-\infty$ to $+\infty$,
we get

\be
n\sim \tau^{-\left(\frac {m\alpha\nu_{||}}{\nu_{||}z_{||}+1}+\frac{(d-m)\alpha\nu_{\perp}}{\nu_{\perp}z_{\perp}
+1)}\right)} \label{non-lin1},
\ee
which reduces to the form (\ref{defect4}) for the linear case $\alpha=1$.

In conclusion, we have proposed that Kibble-Zurek scaling can be studied in
optical lattices of trapped fermions by creating geometries that have
Dirac, semi-Dirac or Quadratic Band Crossings. These systems are 
typically interpenetrating lattices, which can be controlled independently.
Thus, one can create a staggered on-site potential that leads to a gap
in the spectra. As this staggered potential is varied through zero, one
crosses a quantum critical point and even a slow quenching would lead
to the creation of a certain density of defects. We have calculated
the density of such defects for the gapped Graphene Hamiltonian and
have generalized the Kibble-Zurek results to anisotropic Quantum Critical points
that arise with semi-Dirac spectra.

We thank Satyajit Banerjee, Debashish Chowdhury, Warren Pickett
and Diptiman Sen for helpful comments and discussions.


\end{document}